\documentstyle[aps]{revtex}
\def\ie{{\em i.e., }}
\def\eg{{\em e.g., }}

\newcommand{\bc}{\begin{center}}
\newcommand{\ec}{\end{center}}
\newcommand{\be}{\begin{equation}}
\newcommand{\ee}{\end{equation}}
\newcommand{\bq}{\begin{eqnarray}}
\newcommand{\eq}{\end{eqnarray}}
\newcommand{\grad}{\vec{\nabla}}
\newcommand{\op}{order parameter }
\newcommand{\lqs}{local quasi-stationarity }

\begin{document}
\noindent

\vspace*{.4in}

\begin{center}
{\Large \bf Non-equilibrium interface equations:
An application to thermo-capillary motion in binary 
systems}

\vspace*{.4in}
Ravi Bhagavatula$^{\dagger}$, David Jasnow$^{\dagger}$ and T. Ohta$^{*}$ \\
{\sl${}^{\dagger}$ Department of Physics and Astronomy, 
University of Pittsburgh }\\
{\sl Pittsburgh, PA 15260}\\
{\sl${}^{*}$ Department of Physics, Ochanomizu University, Tokyo JAPAN 112 }

\vspace*{0.3in}
{\sl  \today}

\end{center}

\vspace{0.4in}

\begin{abstract}

\noindent
Interface equations are derived for both binary diffusive and binary
fluid systems subjected to non-equilibrium conditions, starting from
the coarse-grained (mesoscopic) models. The equations are used to
describe thermo-capillary motion of a droplet in both purely diffusive
and fluid cases, and the results are compared with numerical
simulations.  A mesoscopic chemical potential shift, 
owing to the temperature gradient, and associated
mesoscopic corrections involved in droplet motion are elucidated.\\[1ex]
\\[3ex]
keywords: interface equations, thermocapillary motion, capillarity

\end{abstract}

\pagebreak

\section{Introduction}
  The study of interfacial dynamics in multi-phase systems has been a
topic of considerable interest in recent years. A wide range of
phenomena such as solidification, viscous fingering,
droplet migration, spinodal decomposition and fracture
all involve, in one way or another, a description of
interfacial motion. Often analysis of these and related phenomena
is performed at the macroscopic level, using, for example, hydrodynamic
equations supplemented by appropriate phenomenological
boundary conditions. Macroscopic {\em interface equations} are 
exceptionally useful both conceptually and computationally
in that bulk degrees of freedom are eliminated;
computationally the cost is the required
careful tracking of the interface position and shape. Some examples
of their application are included in Ref.~\cite{int_eq_refs} and
citations therein.

The macroscopic approach, while valuable in its own domain, has natural
limitations. The most obvious lie in situations in which the
scale of structures is not sufficiently larger than the physical
interfacial thickness, which can grow to several molecular length scales
not too far from criticality, for example.
Phenomena such as coalescence ultimately
run macroscopic approaches to their limits, although ad hoc but reasonable
methods are often available for resolving singularities having to do with
infinitely sharp interfaces. In addition, one may wonder
whether macroscopic boundary conditions may require correction in 
situations which strain the assumptions. 

Derivations 
of macroscopic interface equations from more microscopic starting 
points can be extremely useful in illuminating just what assumptions are 
necessary and the potential limitations. Indeed considerable progress
has been made in this regard beginning from a coarse-grained
or mesoscopic level.~\cite{ohta_kawasaki}
However, the
interfacial response to non-equilibrium perturbations  or inhomogeneities
such as an imposed small thermal gradient, can pose interesting 
challenges. In
this paper, we develop interface equations starting from
mesoscopic (coarse-grained) models for both purely diffusive 
binary systems and binary systems with
hydrodynamic interactions, subjected to small non-equilibrium
perturbations. Using this approach, we investigate 
as an example, thermo-capillary driven
motion of a droplet in both binary diffusive and binary fluid systems
in an imposed temperature gradient. Note that motion of a droplet in
the hydrodynamic case has been well explored by the use of macroscopic
techniques~\cite{Lamb,Young,ss0}, while diffusion driven thermo-capillary
motion has not been received much attention in the
literature~\cite{Bhagavatula-Jasnow}.  Thus, to the best of our
knowledge, our modeling approach provides a first detailed comparison
for the droplet motion in these two distinctly different
cases. Furthermore, in addition to reproducing the results consistent
with macroscopic hydrodynamics, the coarse-grained modeling approach
also provides a means to investigate mesoscopic corrections associated
with finite interfacial thickness. This feature may prove useful in
understanding interfacial dynamics at mesoscopic length scales.

The remainder of this paper is laid out as follows. In the next 
Section we introduce the notation and 
briefly review the coarse-grained models we will 
consider along with the relevant equilibrium features. 
Section~\ref{sec:noneq}
deals with the introduction of the types of non-equilibrium considered 
here and with the effect on a single flat interface. Section~\ref{sec:int-eqs}
contains the derivation of the macroscopic interface equations and
mesoscopic corrections in both the fluid and diffusive cases.

\section{Models}
\label{sec:models}
We begin with the d-dimensional coarse-grained
(mesoscopic) binary fluid and binary diffusive models which are
commonly referred to as the Model-H and Model-B in the
literature~\cite{H-H}. These traditional models have been extensively
used to investigate critical fluctuations in binary systems.  However,
the usefulness of such models for regions well below the critical
point is still being established, by, for example,
demonstrating that in particular limits well 
established macroscopic results are reproduced.~\cite{Jasnow-Vinals} 
This remains a focus in this paper.
One advantage of this
type of modeling is that the physics of capillarity is naturally
built-in as opposed to being introduced through boundary conditions in
macroscopic techniques.

\subsection { Hydrodynamic Model} 
The coarse-grained 
model is defined by a conserved order parameter $\phi$ (\eg
concentration of one of the phases) whose evolution is given by
\be
\frac{\partial\phi(\vec r,t)}{\partial t} 
+ \vec{\nabla}\cdot(\phi \vec {v})
  = \nabla^2 \mu(\vec r,t) 
\label{eq:modH}
\ee
where 
$\mu=(\frac{\delta F}{\delta \phi})$ is
the appropriate 
chemical potential given by the functional derivative of the
Helmholtz free energy $F$. The fluid velocity $\vec v(\vec r, t)$
satisfies a modified Navier-Stokes equation,\cite{H-H,Jasnow-Vinals}
\be
\rho \frac{\partial \vec{v}}{\partial t}= \eta \nabla^2\vec {v} -\vec{\nabla}
 P + \mu \vec{\nabla}\phi
\label{eq:NS}
\ee
where $P$ is the pressure, $\eta$ and $\rho$ are viscosity and density
respectively, which are assumed to be fixed and equal for both
phases for simplicity. 
Equations~(\ref{eq:modH}) and (\ref{eq:NS}) along with the incompressibility
condition, $\vec{\nabla}\cdot \vec{v} = 0$,
completely specify the hydrodynamic model with the
following three boundary conditions: (B1) $\partial \mu/\partial n =
0$; (B2) $\partial \phi/\partial n =0$ and, for example, 
(B3) $\vec {v} = 0$ on the boundary.
Here $\partial/\partial n$ corresponds to the normal derivative at the
{\em system} boundary. The condition
B1 preserves the global conservation of order parameter $\phi$, 
B2 is the natural boundary condition which demands smoothness
of $\phi$ at the edges, and B3 enforces ``no-slip,'' with
other possibilities easily included.
Note that our interest here is to address small Reynolds numbers
($Re$) and, hence, the nonlinear convective terms have been
dropped in the velocity equation~(\ref{eq:NS}).

\subsection { Diffusive Model}
Model-B is in turn
specified by evolution equation
\be
\frac{\partial\phi(\vec r,t)}{\partial t} = \nabla^2 \mu
\label{eq:modB}
\ee
with boundary conditions B1 and B2 mentioned above.  One may view this
as the diffusion dominated limit of Eq.~(\ref{eq:modH})
where the fluid velocity is
neglected. This is a version of Cahn-Hilliard model for
binary diffusive systems used to study phase separation~\cite{Langer}.

The mean-field level equilibrium properties 
of the two coexisting phases are characterized by the local Helmholtz 
free energy
density which, for systems of interest, we take as a
polynomial function
\be
\label{local-fe}
f= \sum_{n >0} \frac{\alpha_n}{n} \phi^n
\ee
with the coefficients $\alpha_n$ depending smoothly on physical
fields such as the temperature chosen
such that the system is within the
two phase regime. Although it is not essential we will generally
assume a scaling equation of state (see below). 
The equilibrium properties depend on the local
temperature $T$ and other field variables that determine the
$\alpha_n$. For example, taking $\alpha_2=-\tau$, where $\tau > 0$
depending linearly on the temperature [$\tau = (T_c-T)/T_c$ with $T_c$
being the mean field critical temperature], and $\alpha_4 = 1$ with
all other $\alpha_n=0$, we have the usual double-well Landau free
energy, which will serve as our ``standard model.''
The (Helmholtz) free energy functional for this case is given
by
\be
F\{\phi\}= \int \{ \ell^2 \frac{(\nabla
\phi)^2}{2} + f(\phi) \} d^dr =
\int \{ \ell^2 \frac{(\nabla
\phi)^2}{2}-{\tau}\frac{\phi^2}{2} + \frac{\phi^4}{4}\} d^dr, 
\ee
where we assume the mesoscopic length $\ell$ is typically of the 
order of several
molecular spacings some distance below the critical point.  
This length characterizes
spatial variations of the order parameter and consequently sets the
interfacial thickness.  The chemical
potential $\mu$ is given by
\be
\mu(\vec r,t) = -\ell^2 \nabla^2\phi + \frac{\partial f}{\partial \phi} =
-\ell^2 \nabla^2\phi - \tau \phi + \phi^3.
\label{chem-potential}
\ee 
Note that our standard model corresponds to a symmetric
situation with equilibrium phase values
$\phi_{eq}=\pm \sqrt{\tau}$; the
coexistence curve is symmetric and, for example,
response functions and other bulk properties are identical in the 
coexisting phases.  When the system is in equilibrium, a flat interface 
with normal along $x$ has the well-known profile 
(exact within this mean-field level in an infinite
system),
\be 
\phi(x)= \phi_{eq}{\tanh}[\frac{x-x_0}{\sqrt{2}\xi}],
\label{eq:tanh}
\ee 
where $x_0$ is the location of the interface and
$\xi=\ell/\sqrt{\tau}$ is the thermal correlation length (see,
e.g., Ref.~\cite{Langer,jasnow-review} for details of this analysis).  
Note that the
profile takes values $\pm\phi_{eq}$ at $x=\pm \infty$ respectively. In
a finite system this serves as a good approximation for the profile so
long as the system size is much larger than the interfacial thickness
$\xi$. We will assume more generally scaling behavior for the order
parameter profile,
$\phi(x) = \phi_{eq}h(\frac{x-x_0}{\xi}),$ with $h(y) \rightarrow \pm 1$
as $y \rightarrow \pm \infty$. Such behavior will follow from
a scaling equation of state determined by 
$f(\phi)$~\cite{fisk-widom,jasnow-review}, or quite generally
with the identification of $\xi$ with $\ell$. 
The profile given in Eq.~(\ref{eq:tanh}) satisfies Eq.~(\ref{chem-potential})
with constant $\mu$.
Note that $\mu=0$ for a {\em symmetric} system at
two-phase coexistence and for the above interfacial
profile.  Using these features, and a scaling
form for the profile, it is straightforward to 
show~\cite{RW,jasnow-review} that the
surface tension $\sigma$ is given by $\sigma (\tau)
= \ell^2 \int (\frac {\partial \phi}{\partial x})^2 dx = \frac{e_1
\ell^2 \phi_{eq}^2}{\xi}$, where $e_1$ is a non-universal constant
depending on details of the profile $h(x)$ and taking the value 
$2 \sqrt{2}/3$ in the standard $\phi^4$-case.  
Note that the surface tension depends on
the temperature (and ultimately on the {\em local} temperature)
through the $\phi_{eq}$ and $\xi$. Also, note that the
analysis can be extended by choosing the coefficients $\alpha_n$ such
that the macroscopic limit (i.e., formally, $\ell \rightarrow 0$) 
of $\sigma$ exists. Physically the macroscopic limit obtains when
the scale of any structures greatly exceeds the interfacial thickness,
which is of order $\xi \sim \ell$.
Specifically, for the above $\phi^4$-symmetric model, the choice
$\alpha_2 = \tau_0/\ell^2$ and $\alpha_4= u_0/\ell^2$ (with $u_0 =1$ for
simplicity) ensures the finiteness of $\sigma$ as well as the
bulk order parameter values in the macroscopic
or sharp interface limit, $\ell \rightarrow 0$.
For specificity we will use the symmetric $\phi^4$ model described above.
The first equality in Eq.~(\ref{chem-potential}) 
is general; we will use it and indicate below which features are 
model dependent.
\section{Introducing driving terms}
\label{sec:noneq}
When the $\alpha_n$
are spatially uniform and chosen such that the system is on the
bulk phase boundary, the system can evolve to a 
simple equilibrium state with a single homogeneous phase or
a two-phase equilibrium with flat interface and volume fractions
chosen according to the overall order parameter.
Considerable progress has been made in the study of
kinetics of the approach to equilibrium and of phase separation in
such cases; see, e.g.,Refs.~\cite{Langer,phase_sep_kinetics}.  
Macroscopic interface equations have been
derived from the coarse grained level in order to 
to explore the physics of phase ordering 
kinetics~\cite{ohta_kawasaki,Ohta}.
However, when one or more of the $\alpha_n$ has a small spatial variation, 
the situation is somewhat different.

In such a situation, the system may be able to reach an inhomogeneous 
equilibrium state depending on the nature of the variation and the
boundary conditions. For example, in a closed system, a fluid in a
gravitational field reaches an equilibrium state with a 
spatially varying 
density. However, in a sufficiently large system, such an 
inhomogeneity can yield interesting dynamical evolution long before the
walls come into play.~\cite{oono-kitahara-jasnow} An example which
produces quasi-stationary states is that of capillarity driven
droplet motion in a temperature gradient.~\cite{Jasnow-Vinals}. Here,
in the simplest case, a droplet drifts toward the `hot' side with 
approximately steady state motion until reaching the
vicinity of the system boundary.
In this approximate steady state, the system is 
spatially non-uniform, but close to local equilibrium everywhere.
In such cases, the interfacial dynamics are influenced by
the spatial variation in the surface tension resulting from the spatial
dependence of the $\alpha_n$. Our interest here is to address this
spatially inhomogeneous
situation and present an approach for deriving 
reduced equations that describe the interfacial dynamics.

For specificity we consider the following spatial dependence in the
standard
example mentioned above: $\tau(x)= -\alpha_2(x) = \tau_0 -\beta x >
0$, $\alpha_4=1,$ and all other $\alpha_n=0$ in the free energy
given by
Eq.~(\ref{local-fe}). This situation corresponds to imposing a
temperature gradient on the system along $x$ direction with the two
ends kept at temperatures corresponding to $\tau_0$ and $\tau_0-\beta
L_x$ where $L_x$ is the size of the system along $x$. 
We restrict ourselves to temperatures slowly varying over the
correlation length so that $\beta \xi << 1$. This is easily
satisfied experimentally, and our analysis remains valid even
for temperature gradients large by experimental 
standards.~\cite{Jasnow-Vinals}
Furthermore we consider only the situation in which the entire
system is in the two-phase region.
Thermocapillary
phenomena associated with such type of thermal gradients have been
numerically investigated recently in Refs.~\cite{Jasnow-Vinals,Llambias,%
Alt-Pawlow,Bhagavatula-Jasnow}. Note that the spatial variation
and hence, the temperature field, is
fixed,  corresponding to  large
thermal conductivity of the medium. Generalizations to treating
the temperature field as an active dynamical variable are possible 
and will be treated elsewhere.

Before considering the evolution of a complicated interface, it is
useful to examine the properties of a stationary flat interface with
its normal along $x$ in a thermal gradient. In such steady state,
which exists for the case of conserved order parameter,
$\partial \phi/\partial t =0$ and $v=0$. Hence following from
Eq.~(\ref{eq:modH}), $\mu$
should be a constant since $\partial \mu/\partial n =0$ at the
system
boundaries. However, the steady state profile $\phi(x)$ 
(consistent with a non-vanishing, constant value of $\mu$)
is a combination of two ramps as shown
from direct simulations of the coarse grained model in Fig.~1.
Below we will use a local equilibrium approximant
which describes the ramp profile very well.
The location $x_0$ of the interface depends on the initial
order parameter content in the system. It is important to recognize
that for {\em conserved} order parameter the profile of Fig.~1 represents
an equilibrium state, while for nonconserved order parameter
the `kink' quite generally moves with constant velocity
toward the hot side as order parameter
is converted in such a way as to reduce the total free energy.~\cite{takakura} 

For the present (conserved) case, by assuming that the interface 
at $x=0$ is at local
equilibrium corresponding to the temperature $\tau(x=0)$, one can
obtain $\mu$ to first order in the gradient $\beta$ as follows.
First, to simplify the algebra we rewrite $\tau(x) = m(x)^2$ with 
$ m(x) = m_0 (1 - b x )$. To $\cal{O}(\beta)$ there is no change if
we identify $b = \beta/2$ and $m_0=1$.
For simplicity we set $\ell = 1$ and
have $\mu = -\frac{d^2 \phi}{dx^2} - m^2(x)\phi + \phi^3.$
Following Ref.~\cite{takakura} we remove the local equilibrium 
order parameter and make a non-linear transformation
defining $\phi(x) = m(x) \eta(z(x)).$ We set the function $z(x)$ by
requiring that the resulting differential equation for $\eta(z)$
has unity for the coefficient of $d^2 \eta/dz^2$. One finds, then, that
$z(x) = m_0 ( x -b x^2 /2 )$ and
\begin{equation}
\frac{\mu}{m(x)^3} = - \frac{d^2\eta}{dz^2}-\eta + \eta^3 + 
(3bm_0/m^2)\frac{d \eta}{dz}
\end{equation}
This procedure effectively isolates the explicit dependence on the
temperature gradient $b \propto \beta$.
Now one can perform an order-by-order analysis, writing
$\eta = \eta_0(z) + b \eta_1(z) + \dots $ and
$\mu = \mu_0 + b \mu_1  + \dots$.
At ${\cal O}(b^0)$ we have
\begin{equation}
\frac{\mu_0}{m_0^3} = 0 = -\frac{d^2 \eta_0}{dz^2}-\eta_0 + \eta_0^3
\end{equation}
with, as expected, $\eta_0 = \tanh(z/\sqrt{2})$. 
The function $\phi_0 = m(x)\eta_0(z)$
corresponds to the local equilibrium ramp and is an excellent numerical
approximation. In fact, if plotted on the same scale as the simulation
result in Fig.~1, there would be virtually no visible difference.

At ${\cal O}(b)$ we have 
\begin{equation}
\frac{\mu_1}{m_0^3} = {\cal L} \eta_1 + \frac{3}{m_0}\frac{d \phi_0}{dz}
\end{equation}
where the operator ${\cal L} = -d^2/dz^2 -1 + 3 \eta_0^2 $ is recognized
as the fluctuation operator for the $\phi^4$ theory, and
$d\eta_0/dz$  is the so-called `translation mode' (see, e.g.,
Ref.~\cite{jasnow-review}). Multiplying through by $d\eta_0/d z$,
and recalling that ${\cal L} d\eta_0/d z = 0$, one finds the
chemical potential shift
\begin{equation}
 \mu_1 = \sqrt{2}m_0^2 b = \frac{\beta \ell}{\sqrt{2}}.
\label{eq:meso-chem}
\end{equation}
In the last equality we have reintroduced the mesoscopic length $\ell$
and have returned to the original definition of the temperature gradient
$\beta$. 

The mesoscopic shift in chemical potential, $\mu_m = \pm \mu_1,$ 
is proportional to the mesoscopic length
scale $\ell$. Also  note that $\mu_m > 0 (< 0)$ depending on whether the
$\phi >  0  (< 0 )$ phase is near the hot end. Physically
the chemical potential shift (from the coexistence value $\mu=0$)
is proportional to the temperature
difference across the interfacial region, which is generally small.
For the
profile shown in Fig.~1, the chemical potential $\mu$ evaluated from 
Eq.~(\ref{chem-potential}) is negative, as shown in the inset. For the 
$\phi^4$ model, using the parameters of Fig.~1, one finds
$|\mu_m| = 0.004/\sqrt{2} \simeq 0.00283$ in excellent agreement with the
direct simulation.
At the time shown in Fig.~1, the interface hasn't quite 
equilibrated in this diffusive system. Small order parameter
changes in the evolution are emphasized in the chemical potential.
For complete equilibration, global diffusion is required.
Simulations on smaller systems and in
one-dimension reveal, more rapidly, a constant chemical potential.
Recall that the homogeneous
case having $\tau = \tau_0 = \mbox{const}$ has $\mu=0$. 

The above analysis arriving at the
leading behavior of a flat interface, \ie 
$\phi~=~\sqrt{\tau(x)} \tanh(z(x))$  can
be viewed as taking the sharp interface limit in which the fast variations
of the order parameter along the normal at the interface are
considered. 
Note that the slow variation associated with the
temperature field is effectively ignored while integrating though the
interface. Furthermore, if the normal to the interface is tilted by an
angle $\theta$ with respect to the $x-$axis (the direction of temperature
variation) one repeats the above analysis using 
$\phi = m(x' \cos \theta ) \tanh( z(x') )$
where $x'$ is the coordinate in
the normal direction. Everything 
goes through as above with the mesoscopic correction 
becoming $\mu_1 \cos(\theta)$ where $\mu_1$ is given in 
Eq.~(\ref{eq:meso-chem}).
Below the calculation will be slightly modified to
deal with a gently curving interface.

Numerical simulations, such as shown in Fig.~1, confirm
that the models exhibit
this shift, supporting the local equilibrium picture of 
interface in an inhomogeneous, slowly varying temperature field.
As noted, a single two-phase interface, owing to the conservation of 
order parameter, can achieve an equilibrium state in a large, finite system.
A {\em slab} consisting of
a combination of two opposite
interfaces, \ie a kink-antikink, cannot. In fact
the mesoscopic shift in chemical potential alone drives the slab to the
hotter side. This phenomenon will be addressed elsewhere~\cite{BLJ}.  

\section{Interface equations} 
\label{sec:int-eqs}
We now consider a gently curved
interface in the coarse-grained modeling and derive equations for the
evolution of the interface. We restrict ourselves
to two dimensions for simplicity
since appropriate generalizations for higher dimensions can be made
straightforwardly. 

To determine the effect of an applied gradient,
we first assume that locally the interface is flat and
is in local equilibrium
with $x$ variation replaced by variation along the normal direction.  
The second
assumption is that the interfacial motion arises from local
quasi-stationary evolution of the order parameter $\phi(\vec r,t) =
\phi(\vec r - \vec V t)$.  In particular, the evolution of the
interface at any point $s$ is characterized by the normal velocity
$\vec {V}(s) \cdot \hat{n}$, where $\hat n$ is a unit
normal (pointing from the `minus' phase to the `plus' phase). 
This means that order parameter or fluid flow along the
tangential direction at any point on the interface do not contribute
to the evolution of the interface.  The equations for diffusive and
fluid cases differ since the mechanisms for the interfacial motion are
different. Interface equations in the absence of imposed gradients
have been well studied in literature~\cite{Ohta}; our interest here is
to extend the derivations and applicability of such equations 
to inhomogeneous
perturbations which can, for example, drive steady state behavior.

\subsection{Diffusive case} 
Eq.~(\ref{eq:modB}) describing the diffusive dynamics
can be inverted (see, e.g., Ref.~\cite{Ohta}) 
using the assumption of \lqs:
\be 
\int d^d r'G(\vec r-\vec r') \frac{\partial \phi(\vec r',t)}{\partial t}  = 
-\mu (\vec r,t) + \lambda(t), 
\label{Dif-1}
\ee
where $G$ is the Green's function for the Laplacian satisfying
$-\nabla^2G(\vec{r}- \vec{r'})=\delta(\vec{r}- \vec{r'}).$
The function $\lambda$ preserves
the \op conservation and satisfies $\nabla^2\lambda =0$ with
zero-gradient boundary condition as does $\mu$. One may
view $\lambda$ as the change in chemical potential due to the presence
of complicated interfaces in the system. The idea is to derive an
equation for the interfacial motion starting from Eq.~(\ref{Dif-1}) by
integrating the fast variation of $\phi$ along the normal at the
interface. A separation of length scales
allows one to take sharp interface limit in which the 
radius of curvature is much greater than the interface thickness; 
an alternative 
procedure for organizing corrections has been used to derive
interfacial dynamics in a 
driven diffusive system.~\cite{Yeung-dds} These methods allow, however,
a significant variation of temperature over the scale of a droplet.

First, the \lqs assumption allows us to replace $\partial
\phi/\partial t$ by $-\vec {V}\cdot \grad{\phi}$. Since the normal
variation of $\phi$ at the interface dominates, we approximate it
further by $-V_n \partial
\phi/\partial n$ at the interface with $V_n$ being the
velocity along the normal. Second, we multiply Eq.~(\ref{Dif-1})
on both sides by $\partial \phi/\partial n$ and then take the sharp
interface limit. At any point $s$ on the interface, by ignoring the
slow tangential variation of $\phi$ along the interface, we can write
\be 
\mu \partial \phi/\partial n \approx 
 [-\ell^2 \frac{\partial^2 \phi}{\partial^2 n} - \tau (x) \phi + \phi^3] 
  \frac{\partial \phi}{\partial n}
- \ell^2 \kappa(s)(\frac{\partial \phi}{\partial n})^2 \ \ ,
\label{eq:mu-dphi}
\ee
where $\kappa$ is the curvature.
On integration through the interface,
the first term in the above is exactly the equation we have for the
flat interface. Hence in the sharp interface limit the first term
yields the mesoscopic correction, 
$\mu_m (\hat{n} \cdot \hat{x})$, as discussed above, while
the second term gives the surface tension multiplied by  the local 
curvature, $\kappa(s)$.  
With these approximations, the following integral equation results for the
normal velocity at the interface
\be 
\int_I ds'  G(s|s')\Delta(s) 
\Delta(s') V_n(s') = \sigma(s) \kappa(s) 
- \mu_m (\hat{n}\cdot \hat{x}) \Delta(s) -\lambda (t) \Delta(s),
\ee
where $G(s|s')=G(\vec r(s)- \vec r(s'))$,  
$\sigma(s)$ is the surface tension and $s$ and $s'$ are points
parameterized by the arc length along the interface, $I$. In the 
above equation
tangential variations in the miscibility gap $\Delta(s)$ give rise to 
to terms in the interface velocity which are
higher order in the temperature gradient $\beta$. The same result 
could be obtained without pausing to find explicitly the mesocsopic
correction to the chemical potential. Since Eq.~(\ref{Dif-1}) will 
be projected onto the interface by action of $\partial \phi/\partial n$,
one may use Eq.~(\ref{eq:mu-dphi}) with the ``inner approximation''
$\phi \simeq \phi^{eq}_I \tanh ( n/ \sqrt{2}\xi_I )$, where the subscript
$I$ indicates the interface location. (Of course, the form of the 
inner approximation is justified by the systematic analysis leading to
the mesoscopic correction.)
In this paper we restrict
ourselves to leading order in $\beta$; hence we set
$\Delta(s)=\Delta$ and obtain
\be 
\int_I ds'  G(s|s') V_n(s') = \frac{\sigma(s) \kappa(s)}{\Delta^2}
- \frac{(\hat{n}\cdot \hat{x}) \mu_m +\lambda (t)}{\Delta}.
\label{dif-int-equation}
\ee
Then the time-dependent Lagrange multiplier
$\lambda(t)$ is determined as part of the solution
guaranteeing conservation of the order parameter as 
specified, in the macroscopic limit, by the vanishing of the 
surface integral,
\be 
\int_I V_n(s) ds = 0.
\ee
The above two equations specify the evolution of a
complicated interface in the 2D diffusive model, including
the leading mesoscopic correction. They can be used,
for example, to
study the evolution of droplets as well as other structures and
interfacial instabilities.~\cite{one-sided} Such
equations, without the
mesoscopic term, have been derived at the 
macroscopic level and used in the literature to study a variety 
of problems such as interfacial growth in an anisotropic Hele-Shaw 
cell~\cite{Sarkar-Jasnow}. 

The interface equations can be used to study the evolution of
complicated interfacial shapes in general. Here we use them to study
thermocapillary response of a spherical droplet of radius $R$ in
$d$-dimensions to a small thermal gradient. The curvature $\kappa(s)$ at any
point $s$ in this case is simply given by $\kappa(s)= (d-1)/R$, and the
integration over $s'$ in Eq.~(\ref{dif-int-equation}) is carried over a
d-dimensional sphere. Note that in the absence of a gradient the
droplet remains stationary since, $\sigma(s)=\sigma$ and $\mu_m=0$. As
a result $\lambda$ is a constant and takes a value $\lambda =
\lambda_0 = \sigma (d-1)/(R \Delta)$. This is immediately recognized as the
required shift in the chemical potential of the system 
off bulk phase coexistence to maintain a droplet 
of radius $R$. In macroscopic treatments which assume local
equilibrium, this shift is embodied in the the Gibbs-Thomson
relation. The treatment here 
amounts to an alternative derivation of this relation from a 
coarse grained starting point. Other derivations
are contained, for example, in 
Refs.~\cite{Caginalp,Langer}. 

In the presence of a thermal gradient, the surface tension $\sigma$ has
spatial variation which drives the droplet. This is the standard
Marangoni effect, but in a purely diffusive system.
In particular, if we assume a
linear variation, $\sigma(s) \approx \sigma_0 + \sigma_1 \beta R
\cos(\theta)$, where
$\theta$ is the angle of the outward normal to the gradient direction, 
and neglect higher order corrections in $\beta$, 
there is a steady-state
solution for the motion of the droplet.  The velocity of the
droplet can be obtained by solving the integral 
equation~(\ref{dif-int-equation}). In
particular, $V_n = V_0 \cos (\theta)$ is a solution with
$\lambda=\lambda_0$, corresponding to rigid motion of the droplet
along the direction of the gradient. 
Following some of the manipulations in the Appendices of
Ref.~\cite{Ohta}, one expands the Green's function for two points
on a sphere of radius $R$ in spherical harmonics as 
\be
\frac{-1}{4 \pi |\vec{a} - \vec{a'}|} = - \sum_{\l=0}^{\infty}
\sum_{m=-l}^{m=+l} \frac{1}{2l+1}\frac{1}{R}Y_{lm}(\theta,\phi)
Y_{lm}(\theta', \phi')^{*}
\ee
where $\theta,\phi$ locate $\vec{a}$ and similarly for $\vec{a'}$.
Inserting this in the integral equation for the drop velocity
with $V_n = V \cos(\theta)$, which takes the form 
%
%
$$-V \int G(\vec{a}, \vec{a'}) \cos(\theta') R^2 d\Omega'
 =\frac{ -\sigma_1 \beta (d-1) \cos(\theta)}{\Delta^2}
-\frac{\mu_m \cos(\theta)}{\Delta},$$
one may obtain the velocity, $V_0$. The calculation in two dimensions
is similar, and one finds
\be
V_0= C(d) [-\frac{\sigma_1\beta}{\Delta^2 R} + \frac{\mu_1}{\Delta R}]
\label{eq:vel_modB}
\ee
with $C=2$ in $d=2$ and $C=6$ in $d=3$. Note that 
typically $\sigma_1 <0$ 
since surface tension normally decreases with increasing temperature, thus
drives the droplet to hotter side.
It is interesting that the mesoscopic correction
with $\mu_1$ defined in  Eq.~(\ref{eq:meso-chem})
is always positive, indicating that it also drives
the bubble toward the hotter side. 
Note that the dimensionality dependence only appears in
the coefficient $C(d)$, and that the mesoscopic correction does not
modify the scaling with droplet radius.  

As a check one can compare the results of
numerical simulation of the 2D diffusive model with the above analytic
expression. The simulations represent a direct forward integration
of Eq.~(\ref{eq:modB}). 
(A sample plot is shown in
Fig.~5a of Ref.~\cite{Bhagavatula-Jasnow})
Good agreement with the $ R^{-1}$ dependence as predicted
in Eq.~(\ref{eq:vel_modB}) is found 
supporting the validity of the interface equation approach based on
local equilibrium. Including the mesoscopic correction
for $\beta = 0.002$ 
one finds from Eq.~(\ref{eq:vel_modB}) $V_0 = 1.56 \times 10^{-4}$
for $R=18$ while the direct simulation yields $V_0 \simeq 1.87 \times
10^{-4}$, while for $R = 24$, Eq.~(\ref{eq:vel_modB}) yields
 $V_0 = 2.8 \times 10^{-3}$ to be compared with $2.71 \times 10^{-3}$
from direct simulation. It should be noted that the mesoscopic
term in Eq.~(\ref{eq:vel_modB}) is significant. 
We have also been able to observe the mesoscopic
correction effects in the numerical simulations for smaller
droplets. A plot of the chemical potential itself most directly
reveals this, as in Fig.~1.

\subsection {Hydrodynamic interactions} 
The first assumption we make here
is that the diffusive time scale of the order parameter is so large
that it can be neglected compared to the viscous time scale set by the
fluid. This is equivalent to
saying that the interface is being primarily advected by the fluid
and that diffusive effects (necessary for full approach to equilibrium)
act as corrections.
Next, similar to the diffusive case, we appeal to quasi-stationary
motion of the interface.  Note that, one has to take both normal and
tangential velocities of the fluid at the interface into
account. However, the fluid velocity normal to the interface
determines the evolution of the interface as in the diffusive case.

The velocity equation can again be inverted,~\cite{Ohta} 
consistent with the quasi-stationary assumption,
yielding the following equation:
\be
v_{\alpha}(\vec r) = \frac{1}{\eta} \int \Sigma_{\beta} 
T_{\alpha \beta}(\vec r, \vec
r') (\mu \grad \phi)_{\beta} d^d r', 
\label{eq:oseen}
\ee 
where $T_{\alpha \beta}$ is the Oseen tensor, the form of which 
ensures the
divergence free nature of the velocity field. This also implies
that the normal velocity $v_n$ on any closed surface satisfies 
$\int_s v_n ds = 0$, which, since
diffusion is neglected as noted above,
is consistent with the conservation of the order
parameter content enclosed within. 
Note that the pressure 
term drops out in the above equation since it has
been chosen to guarantee divergence free flow. 
In Eq.~(\ref{eq:oseen}), the $\mu \grad \phi$ term is large
mainly at the interface. One can integrate this
term though the interface, similarly to the steps carried out for
diffusive case, and arrive at the following interface equation:
\be 
v_{\alpha}(s) = \frac{1}{\eta} \int \Sigma_{\beta} T_{\alpha
\beta}(s|s') F_{\beta}(s') ds',
\label{fluid-int-equation}
\ee 
where $s, s'$ are any two points on the interface and
$T(s|s')=T(\vec{r}(s)-\vec{r}(s'))$. The force $F_{\beta}(s)$ at the
interface is obtained by integrating the $\mu
\grad \phi$ through the interface.  The normal and tangential forces
$F_n$ and $F_t$ along the unit normals $\hat n$ and $\hat t$ 
at any point $s$ on the interface are given by
\bq
F_n(s) & = & -\kappa(s)\sigma(s) + \mu_m \Delta (\hat{n}\cdot \hat{x})\\ 
F_t(s) & = & \frac{\partial \sigma(s)}{\partial s} + \mu_m \Delta
(\hat{t}\cdot \hat{x}), 
\label{eq:forces}
\eq 
where $\kappa$ is the curvature and $\sigma$ is the surface tension.
Note that in higher dimensions the tangent surface is
multi-dimensional.  Since the gradient is applied along one specific
direction, we can work with a single tangent vector by exploiting the
azimuthal symmetry.  The first terms on the
right hand side in the above two force
equations are macroscopic and are identical to the terms
phenomenologically included by assuming a sharp interface between the
two phases in the standard macroscopic hydrodynamic analysis.~\cite{Young}  
The terms involving $\mu_m$ are mesoscopic corrections, which will 
occur for finite size droplets.
Hence the coarse-grained models
reproduce the macroscopic results exactly. The tangential force
derives macroscopically from
the presence of a tangential variation of the
surface tension, which itself follows in situations such as those
with an imposed thermal gradient.

Now we use the above interface equation (\ref{fluid-int-equation}) to
obtain the velocity of a spherical droplet of radius $R$ in a small
applied thermal gradient $\beta$. Note that in the absence of
gradient, the droplet remains stationary, and the pressure field
satisfies Laplace's law in the model consistent with the macroscopic
expectations. By assuming an approximate
linear variation of the surface tension
with temperature, a point to which we will return below, \ie
$\sigma(s) = \sigma_0 + \beta R
\sigma_1 \cos(\theta)$, we can integrate the interface equation to
obtain the droplet velocity. In the quasi-stationary approximation the
center of mass of the droplet is assumed to move with a constant
velocity $V$ along the gradient in the lab frame, leading to a normal
velocity $V \cos(\theta)$, where $\theta$ is the azimuthal
angle. This is due to the fact that, in the center of mass frame, 
the droplet must be stationary, and, hence, the normal velocity 
must vanish. Even
though the component of the fluid velocity is zero along the normal,
the tangential velocity need not be zero in CM frame. This turns out
to be the case for thermocapillary motion of a droplet in
the hydrodynamic case.

The CM velocity of the droplet V can be obtained by solving the 
integral equation in three dimensions using specifically the Oseen tensor  
$T_{\alpha \beta} (\vec r)= \frac{1}{8\pi r} (\delta_{\alpha\beta} + 
\frac {r_{\alpha} r_\beta}{ r^2})$. 
In three dimensions a useful identity~\cite{Ohta} is
\be
\int d\Omega' \sum_{\alpha,\beta}n^{\alpha}(\Omega)
T^{\alpha\beta}n^{\beta}(\Omega')
Y_{lm}(\Omega') = \frac{1}{R} \frac{2l (l+1)}{(2l-1)(2l+1)(2l+3)}
Y_{lm}(\Omega)
\ee
and for $l=1,m=0$ the integral equals $4 \cos(\theta) / 15 R $.
Using such results one finds for the CM velocity,
\be 
 V = -\frac{2}{15 \eta} (\sigma_1 \beta R) + \frac{2}{3 \eta} 
(\mu_1 \Delta R).
\label{eq:3dvel}
\ee
The first term is identical to the solution obtained via
macroscopic analysis of the Navier-Stokes equation~\cite{Young}
for the special case of two fluids with identical fluid and
thermal properties. The second term represents the mesoscopic 
correction for the droplet velocity,
where $\mu_1$ was evaluated in Eq.~(\ref{eq:meso-chem}).
In two dimensions the Oseen tensor is
logarithmic, and one needs to introduce a screening length $a$
presumably set by the droplet size. Explicitly, $T_{\alpha \beta} (\vec r)=
-\frac{1}{4\pi} [\delta_{\alpha\beta} log (r/a) - \frac {r_{\alpha}
r_\beta}{ r^2})$.  With the  simplest assumption $a=R$, similar
manipulations yield the 
center of mass velocity
of the droplet as
\be 
V = -\frac{1}{8 \eta} (\sigma_1 \beta R) + \frac{1}{2 \eta} (\mu_1 \Delta R).  
\label{eq:2dvel}
\ee
Interestingly, the macroscopic part, \ie the first term on the
right, does not depend on the screening length, $a$.
For general screening length, the mesoscopic, second, term is
multiplied simply by the factor $(1 + \ln(R/a))$.
For a suspension of many droplets, it is reasonable to expect that
$a \sim R$, but for a single droplet, the choice remains
problematic.
As pointed out elsewhere~\cite{Jasnow-Vinals} the $R$ dependence of the
macroscopic term is expected to be dimensionality independent.
In the coarse grained model we are considering here, the surface tension
decreases with increasing temperature, \ie , $\sigma_1 < 0$.  Hence
the droplet is driven along the direction of the gradient towards the
warmer side.  Note that the mesoscopic term also drives the droplet to
the warmer side with the scaling of velocity with radius of the
droplet being unchanged; the total proportionality
constant is shifted, however. 

As a check on our ability to analyze the motion of a droplet in the
mesoscopic model, we have numerically investigated the 
average velocity of the droplet and the velocity of
the fluid at the interface of the droplet in the
two-dimensional case. 
The simulation methodology and data arise from
calculations described in
Ref.~\cite{Jasnow-Vinals}, in which Eqs.~(\ref{eq:modH}) and (\ref{eq:NS})
were forward integrated. 
The macroscopic term in Eq.~(\ref{eq:2dvel})
yields $ V \simeq 0.012$ for the $\phi^4$ model for droplet size
$R = 24$ ( and operating conditions $\tau_0 = 0.5, \beta = 0.001, 
1/\eta = 5.96 $) which correspond to a simulation in that reference.
The numerical agreement is excellent; Fig.~3 of Ref.~\cite{Jasnow-Vinals}
shows $V \simeq  0.0005 R \simeq  0.012$. This agreement does not address the
ambiguity of the screening term in the mesoscopic correction in
two dimensions.
In Fig.~2, we show the velocity $v_x$ and
$v_y$ at the interface of the droplet as a function of the azimuthal
angle $\theta$. One can clearly see the variation consistent with a
non-zero tangential fluid velocity at the interface. 
One can show that
the macroscopic part of the tangential velocity is 
$V \sin(\theta)$ leading to $v_x = V
\cos(2\theta)$ and $v_y = -V \sin(2\theta)$, confirming non-solid body
motion of the droplet in the fluid case.~\cite{footnote}
This behavior agrees with
macroscopic analysis based completely on the steady solution of the
Navier-Stokes equation (\ie, by redoing the analysis of
Refs.~\cite{Lamb,Young} for comparison with the two-dimensional 
simulations.)
The order of magnitude of $V$ obtained in
the simulations is in agreement with the analytic estimate
in Eq.~(\ref{eq:2dvel}) above. 
Additional
quantitative investigations are left for future explorations. 
Appropriate checks should involve three-dimensional calculations
to avoid the ambiguity of the screening length in the mesoscopic
correction.
The presence of the mesoscopic correction can be verified using the chemical
potential profile which shows linearity inside the droplet along
the gradient direction (\ie $ \nabla^2 \mu$ is small consistent
with the assumptions), corresponding to the mesoscopic shift in the
chemical potential at the interface. 

The inclusion of a body force such as gravity into
the interface equations for two fluids having different densities has
been
discussed earlier in Ref.~\cite{Ohta}, but without a thermal gradient.
In the presence of a thermal gradient, it can be checked that the
normal force $F_n$ at the interface gets appropriately modified (a
term $g (\Delta\rho) \cos(\theta)$ with $g$ the acceleration due
to gravity and $(\Delta\rho)$,
the density difference enters), leading to an additional
macroscopic term identical to the result of Young 
{\em et al.}~\cite{Young}
For example,
in three dimensions 
the gravity coupling yields the following CM velocity for the
droplet:
\be 
 V = -\frac{2}{15 \eta} (\sigma_1 \beta R) + 
 \frac{4}{15 \eta} (C R^2) + 
\frac{2}{3\eta} (\mu_1 \Delta R), 
\ee
with $C=g (\Delta\rho) $.  
Note the different scaling of the macroscopic terms with droplet
radius, $R$, indicating the interplay between a bulk body
force and a force concentrated at the interface. With proper
orientation and strength of the gradient relative to the gravitational
force, the macroscopic terms can be tuned to cancel, leaving a 
small systematic drift toward the hot side due to the mesoscopic
correction.
Phase asymmetry (\ie owing to a free energy
without $\phi \rightarrow -\phi$ symmetry)
also couples to a temperature gradient in a way similar to gravity
modifying only the the normal force on the
interface~\cite{Bhagavatula-Jasnow}. 

\section{Concluding remarks} 
In this work we have investigated macroscopic equations
for interfacial dynamics arising from two coarse-grained 
dynamical models. These equations agree with those derived
from purely macroscopic considerations, which provides additional
evidence for the utility of these coarse-grained models
even away from the critical regime; furthermore the details of the
analysis indicate clearly the assumptions required and potential
sources of corrections.

More specifically, we derived interface equations
starting from coarse-grained descriptions of binary systems
whose dynamics are dominated by diffusion (Model-B) 
and by hydrodynamic interactions (Model-H)
using a local equilibrium
description. We allow for spatially varying potentials such as
an imposed thermal gradient, as long as the spatial variation
is slow on the scale of the interfacial width; such variation
may nonetheless be macroscopically large so as to be significant
over the scale of the relevant structures. 
The equations are shown to describe thermocapillary driven 
motion of a droplet quite well, yielding
complete agreement with macroscopically derived results
in the appropriate limit in which structures
are large with respect to the interfacial width and 
the gradient is sufficiently weak to neglect all but 
the linear temperature coefficient of the surface tension. 

Clearly the interface equations derived are more general.
The coarse-grained models naturally include non-linear
dependence of the surface tension on the temperature.
We also show how mesoscopic effects
associated with the finite thickness of the interface
do not shrink as the radius increases.
The mesoscopic
effects alone can play a vital role in driving 
thermocapillary migration in
situations in which the curvature effects are less important 
(\eg  motion of
a slab, or kink-antikink pair in a two phase system). 
We have added additional circumstantial evidence for the 
applicability of
coarse-grained free energetics and coarse-grained
hydrodynamics even away from criticality, for which the modeling
was originally designed. 
The local equilibrium or ``local steady
state''~\cite{Yeung-dds}  analysis and the interface equation approach are
sufficiently general to allow application
to a variety of other driven systems.

\noindent {\bf Acknowledgments:}
RB is grateful for support under grant NAG3-1403 from NASA Microgravity
Science and Applications Division. 
DJ thanks the Japan Society for the Promotion of Science for
partial support of this work as well as the NSF under DMR92-17935.
We are grateful to the NCCS at the
NASA Goddard Space Flight Center for providing supercomputer time on
the Cray-C98.  Computer time on Cray T3D from the Pittsburgh
Supercomputing Center is also acknowledged. Thanks go to Prof.
D. Boyanovsky for his interest and helpful suggestions.

\newpage 

\begin{center}
{\bf Figure Captions}
\end{center}

\noindent {\bf Fig.1:} The order parameter profile
resulting from direct numerical simulation in
two dimensions is shown for 
$\beta= 0.004$, $\tau_0 = 1$ and $\ell=1$ in units of thermal
correlation length. One can think of the profile as a combination of
two ramps having slopes proportional to $\beta$. A system of length 100
is used. The inset shows the chemical potential evaluated
from Eq.~(\ref{chem-potential}) at the corresponding time during the
approach to equilibrium.

\noindent {\bf Fig.2:} Simulation of droplet motion with hydrodynamic 
coupling.
(a) The $X-$component of the velocity at the droplet 
interface (i.e., component 
along the direction of the gradient) is shown as a
function of the azimuthal angle ($\theta$).  The solid line
indicates, for comparison,
$ cos(2\theta)$ behavior. (b) The $Y$-component is shown. 
For comparison the solid line shows
$- \sin (2\theta)$ behavior. 
These indicate
non-rigid motion for the droplet. The parameters used are $\beta = 0.001$, 
$R = 24$, $\tau_0 =0.5$. A $200 \times 200$ system was simulated
in units of $\ell$.

\input epsf
 
\begin{figure}[h]
\centering
\epsfysize=5.0in  
\hspace*{0in}  
\epsffile{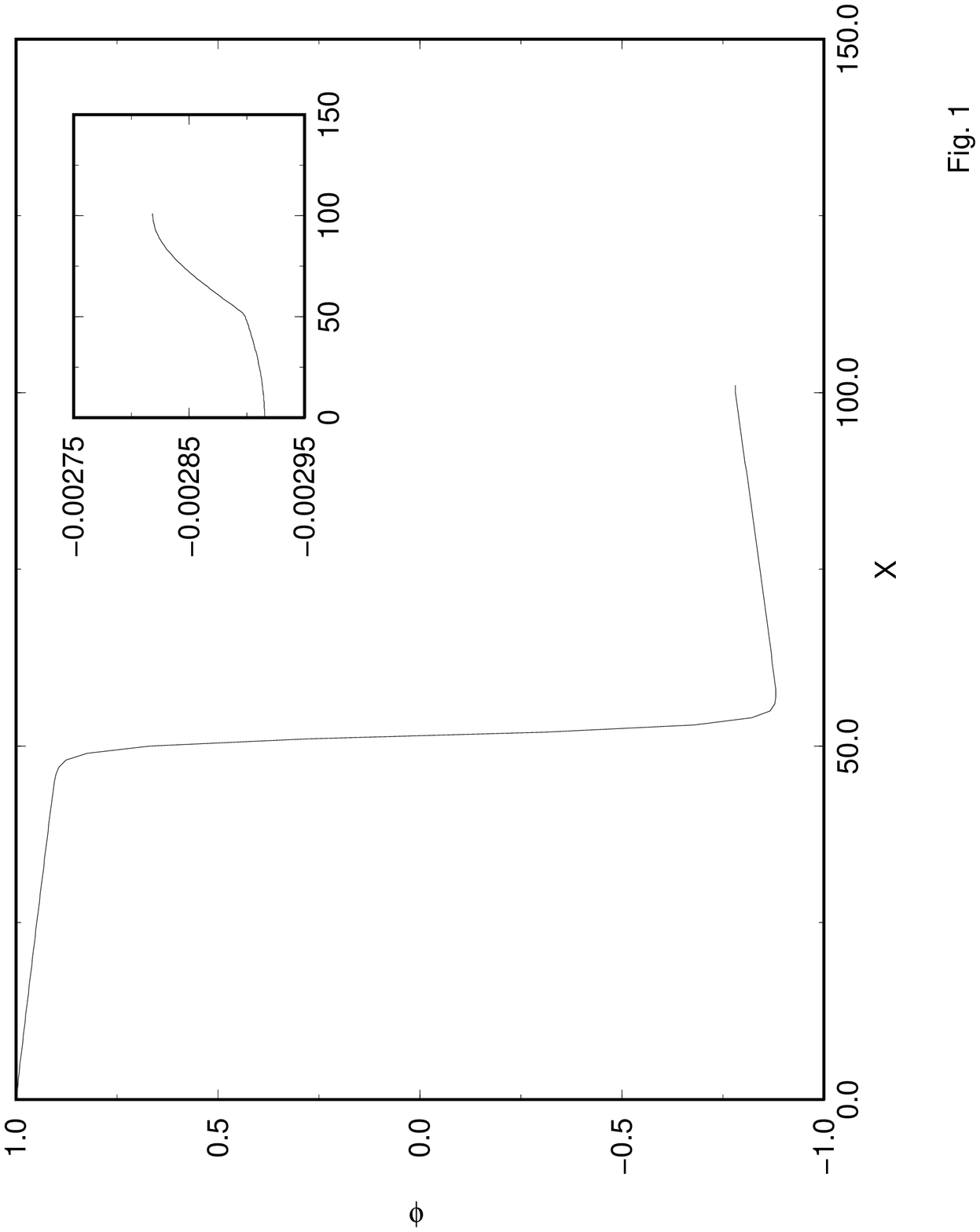} 
\caption{}
\label{fig:engulfing} 
\end{figure}

\begin{figure}[p]
\centering
\begin{minipage}[t]{4in}
\epsfysize=4.0in
\hspace*{0in}
\epsffile{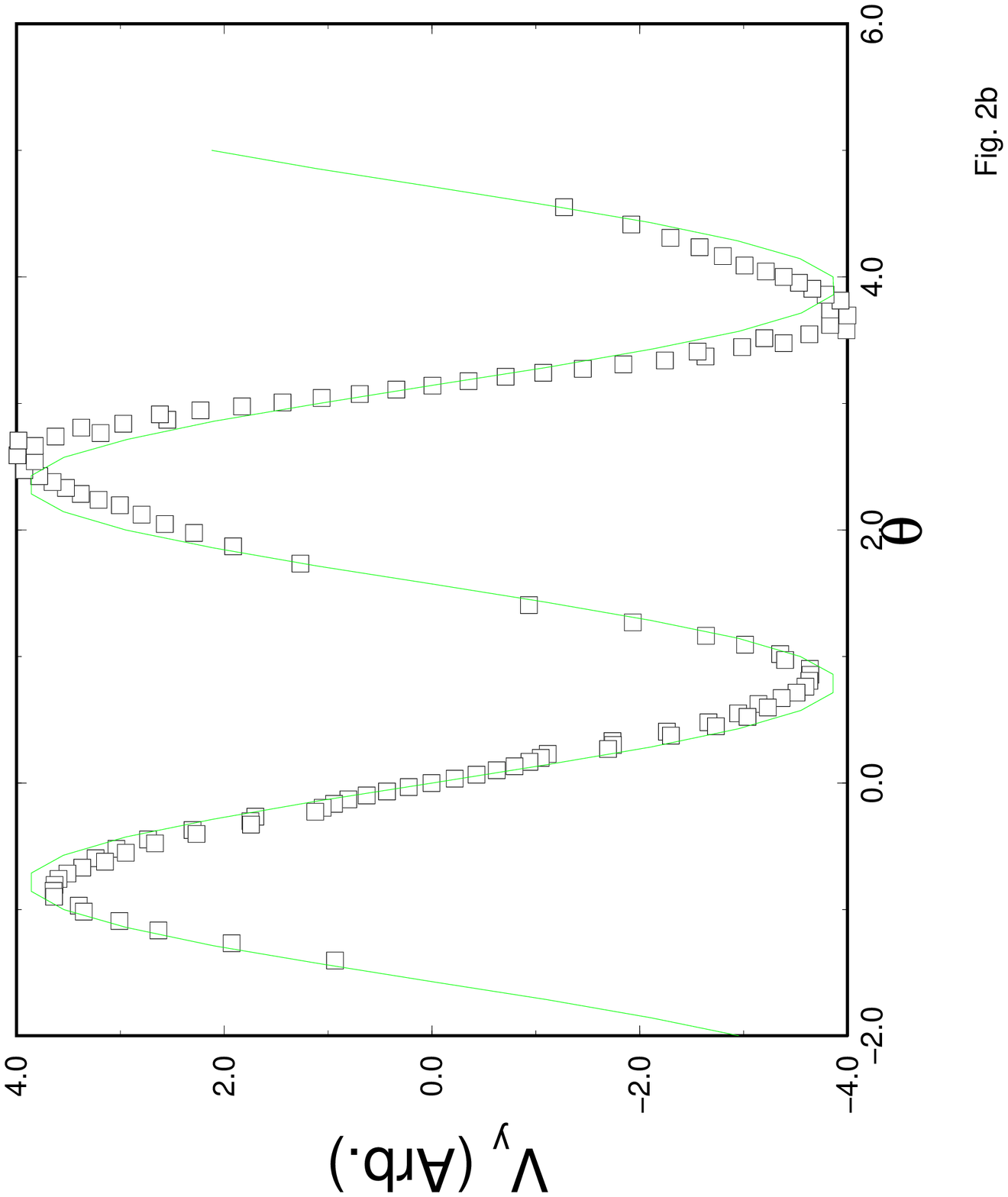} 
\end{minipage}
\hfil
\begin{minipage}[t]{4in}
\epsfysize=4.0in
\hspace*{0in}
\epsffile{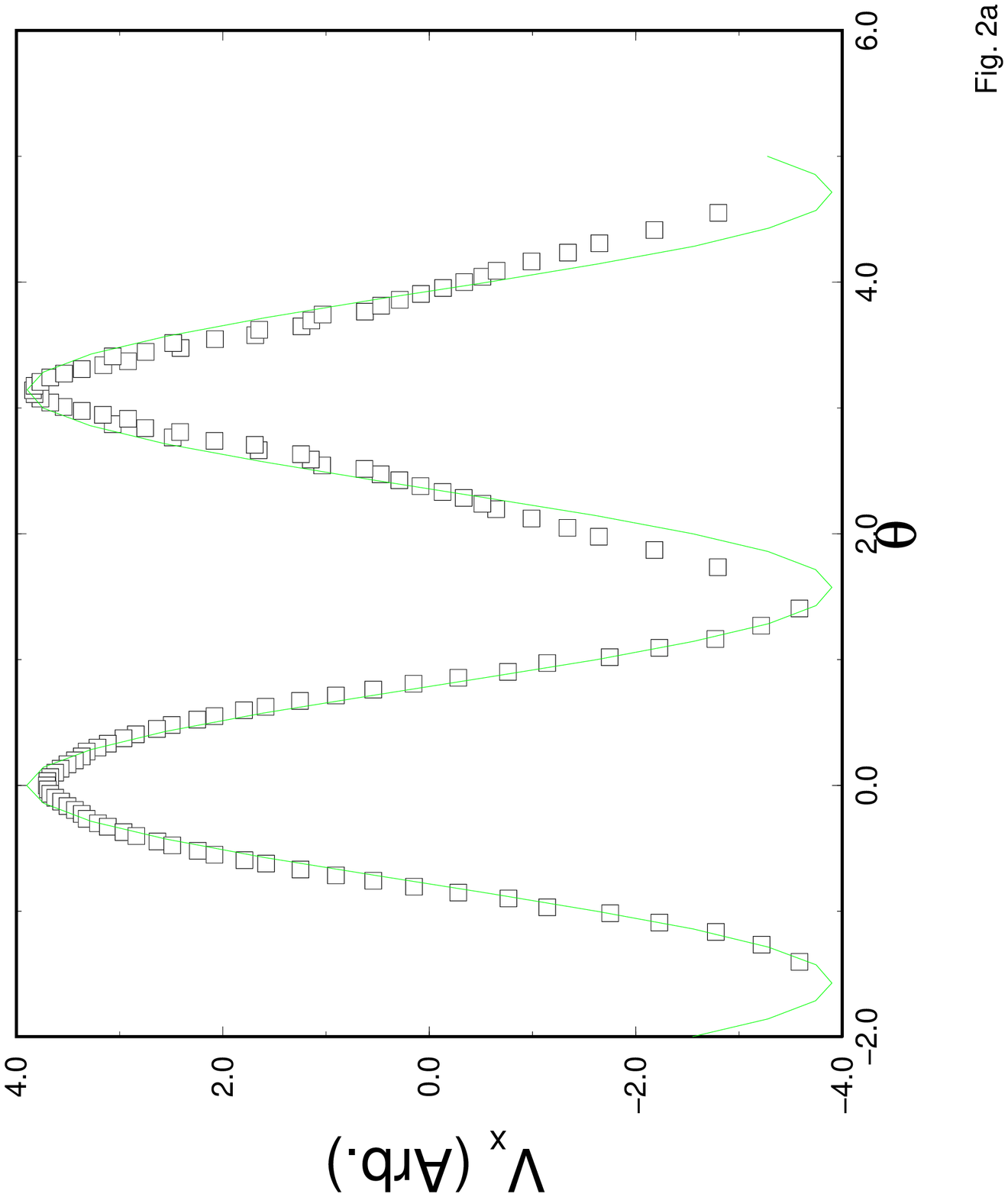}
\end{minipage}
\caption{}
\end{figure}


\begin{thebibliography}{99}
\bibitem{footnote}
If the droplet were moving steadily as a solid object, the
interface velocity would vanish in the CM frame, \ie, the rest
frame of the droplet. Then 
would have $v_n = V \cos(\theta)$ and
$v_t = -V \sin(\theta)$.
\end{thebibliography}
\end{document}